# DEEP LEARNING CLASSIFICATION OF CHEST X-RAY IMAGES

Mohammad S. Majdi[1,3], Khalil N. Salman[2], Michael F. Morris[4], Nirav C. Merchant[3], Jeffrey J. Rodriguez[1]

{[1]Dept. of Electrical and Computer Engineering, [2]Dept. of Medical Imaging, [3]Data Science Institute}. University of Arizona, Tucson, AZ, [4]Dept. of Radiology, University of Arizona, Phoenix, AZ.

Email: {mohammadsmajdi, nirav, jjrodrig}@email.arizona.edu, ksalman@radiology.arizona.edu, Michael.Morris@bannerhealth.com

## ABSTRACT

We propose a deep learning based method for classification of commonly occurring pathologies in chest X-ray images. The vast number of publicly available chest X-ray images provides the data necessary for successfully employing deep learning methodologies to reduce the misdiagnosis of thoracic diseases. We applied our method to the classification of two example pathologies, pulmonary nodules and cardiomegaly, and we compared the performance of our method to three existing methods. The results show an improvement in AUC for detection of nodules and cardiomegaly compared to the existing methods.

*Index Terms* — chest X-ray, deep learning, classification, pulmonary nodule, cardiomegaly.

## 1. INTRODUCTION

In recent years, due to the abundance of chest X-ray (CXR) images, deep learning (DL) has gained wide popularity in the analysis of radiographic images and is anticipated to help radiologists in the context of disease detection and management [1]–[3]. Valley fever (coccidioidomycosis) is a fungal infection that has been endemic in the southwestern areas of the United States for hundreds of years [4]. Valley fever causes pneumonia-like symptoms and may be disseminated leading to other systemic infections that may manifest in meningitis, leading to cognitive impairment, paralysis or death [5]. Valley fever may also manifest as pulmonary nodules which may be mistaken for lung cancer. In the past decade, the infection incidents of this disease have skyrocketed, reaching 22 500 cases in 2011, up from 2265 in 1998 [6], [7]. Furthermore, the difficulty in monitoring and diagnosing valley fever has led to misdiagnoses; thus, the statistics may not fully characterize the magnitude of this disease [8]. Following the detection of a pulmonary nodule, depending on the size, diagnostic differentiation methods available are limited and primarily involve sequential imaging with CXR, CT, PET-CT or invasive tissue sampling. An automated method can provide a second opinion and thus mitigate the problem of misdiagnosis. Furthermore, the existence of a fully automated method can help with diagnosis in less developed regions that don't have access to an experienced physician.

X-rays are one of the most commonly used and publicly available radiological modalities. Recently, multiple groups have collected a vast number of CXR images, creating large datasets available to machine learning researchers. CheXpert [9] consists of 224 316 chest radiographs of 65 240 patients, labeled for the presence of the 14 most prevalent diseases in clinical reports as shown in Fig. 1. PadChest [10] consists of 160 000 chest radiographs obtained from 67 000 patients at San Juan Hospital (Spain) from 2009 to 2017. Finally, the NIH chest X-ray dataset consists of 100 000 chest radiographs.

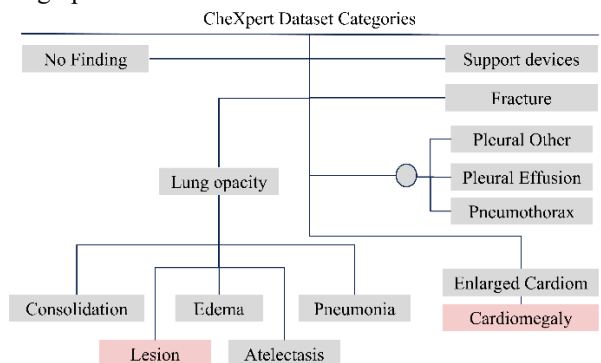

Figure 1: CheXpert dataset disease classification. The two pathologies we studied are colored in red.

Wang [11] uses a unified weakly supervised multi-label image classification and disease localization framework to detect the commonly occurring thoracic diseases. Irvin [9] experimented with various existing deep architectures and approaches to treating the uncertainty in the manual labels to classify the 12 commonly occurring pathologies in their chest X-ray dataset. Various studies have explored the applicability of analyzing such datasets using deep learning to help with radiological imaging diagnosis. For example, multiple studies have used machine learning for automated diagnosis of tuberculosis from radiographic images [12]–[15].

Taylor [16] used deep learning to detect pneumothorax in CXR images, reporting the receiver operating characteristic (ROC) area under the curve (AUC) as 0.94–0.96. Yao [17] used long short-term memory (LSTM) to leverage interdependencies among target labels in CXR images. Zech [18] investigated the generalizability of deep learning based models on images collected at different sites (hospitals) to detect pneumonia and reported an AUC of 0.93-0.94 when the model was trained on images obtained from the same site, and a significantly lower AUC of 0.75-0.89 when trained on images from other sites.

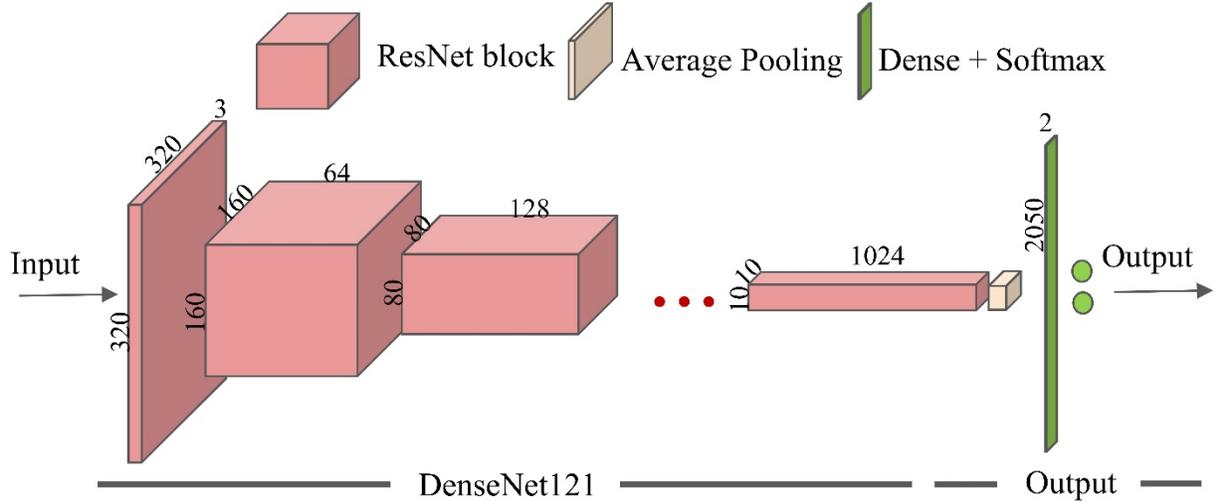

Figure 2: Proposed framework – Densenet121 [21] plus global average pooling and softmax layers.

Pulmonary nodules are spots smaller than 3 cm [19] that appear in medical chest images. Pulmonary nodules usually cause no symptoms and are discovered when a patient has gone through a CXR or CT scan for another matter. The presence of a nodule may indicate abscess, tuberculosis, pneumonia, or coccidioidomycosis (valley fever) [20]. In this study, we have developed a deep learning based method for the detection of pulmonary nodules (referred to as lesions in the CheXpert [5] dataset) to help with the diagnosis of thoracic diseases such as valley fever. Even though Irvin [9] uses a similar architecture to ours, what distinguishes our work from theirs is the use of a weighted loss function to account for the huge class imbalance in the CheXpert dataset for pulmonary nodules. Further, Irvin [9] only reports AUC for five pathologies not including pulmonary nodules, which could be due to this class imbalance. In order to compare the performance of our proposed method against Irvin's, we tested our method on another pathology, cardiomegaly, which was randomly selected from the five pathologies that Irvin analyzed to investigate the effect of using our weighted loss function against their reported AUC.

## 2. METHODS

After exploring different architectures, we chose to use a dense convolutional network (DenseNet) architecture [21] for the main body of the proposed method. DenseNet connects the output of each layer of the network to all its subsequent layers, so that each layer of the network receives the outputs of all preceding layers as inputs. This has helped mitigate the problem of vanishing gradient and substantially reduced the amount of memory required for training very large datasets. In order to improve the convergence rate, all network parameters were initialized with parameters of a similar network trained on the ImageNet dataset [22]. Finally, in order to adapt this architecture to our problem, the last fully connected layer was removed, and each feature map in the previous layer (1024 feature maps of size 10×10) is averaged and concatenated using global average pooling to create a feature vector of size 1024. Finally, another dense layer (with a size of 2×1) followed by a 1×1 convolution with a softmax activation function is used to map these feature vectors into the desired classes, where the output of the network will be two probability maps showing the likelihood of the input image belonging to each class (positive or negative). The final architecture is shown in Fig. 2.

All input images are normalized to have zero mean and unit standard deviation. Due to the significant class imbalance, a weighted loss function is used, where each class's weight is computed as

$$w_p = \frac{N_P + N_n}{N_P} \quad , \quad w_n = \frac{N_P + N_n}{N_n} \qquad (1)$$

where $N_P$ and $N_n$ are the number of positive and negative cases in the training dataset.

A weighted binary cross-entropy [29] is used as the loss function:

$$L1 = -\sum_{i=0}^{C-1} w_i\, q_i \log(p_i) \qquad (1)$$

where $C = 2$ is the number of classes (positive/presence and negative/absence of the pathology), $p_i = \exp(\hat{y}_i)/\sum_{k=0}^{C-1}\exp(\hat{y}_k)$ is the probability of the input image belonging to class $i$ after applying the softmax function to the network's final score $\hat{y}_i$, $w_i$ represents the weight for class $i$, and $q_i$ is the ground truth for class $i$.

Both networks (for pulmonary nodules and cardiomegaly) are trained using the Adam optimizer [23] with 15 epochs and a batch size of 20. The learning rate is set to 0.0001 throughout the training. In order to avoid overfitting and preserve the best sets of parameters, the weights for the epoch with least validation loss are used as the final network parameters. The final network output's probability maps are converted to binary values using the threshold that corresponds to the highest average F1 score on the validation set across each pathology. The network is implemented in Python and Keras [24] with a TensorFlow backend using an NVIDIA P100 GPU with 16 GB GDDR5 RAM. The required training times for the lung nodule (lesion) and cardiomegaly networks were 7 and 19 hours, respectively. The number of trainable and non-trainable parameters in both networks were 6 955 906, and 83 648, respectively.

The CheXpert dataset consisting of 224 316 chest radiographs of 65 240 patients is used in this study. All labels are extracted by applying a universal dependency parser [28] to the clinical reports corresponding to each patient. Each report is split and tokenized into sentences using NLTK [25]. Each sentence is parsed using the BLLIP parser [26]. The universal dependency graph of each sentence is computed using Stanford CoreNLP [27]. Finally, the extracted keywords are matched against the manually extracted keywords by experienced radiologists.

All front-view images of cases with lesions are extracted from the CheXpert dataset, resulting in 1270 and 9189 cases with positive and negative labels, respectively. Due to the very few cases of patients with lesions in the validation folder of the CheXpert dataset (1 case out of 200), the provided training folder is randomly divided into training (64%), validation (16%) and testing (20%). Further, to reduce the amount of required memory, all input images are downsampled to 320 × 320. The distribution of training, validation and testing data is shown in Table 1.

Table 1: NUMBER OF CASES FOR EACH PATHOLOGY

|  | Training + Validation | | Testing | |
| --- | --- | --- | --- | --- |
|  | Pos | Neg | Pos | Neg |
| Pulmonary nodule | 935 | 7430 | 335 | 1756 |
| Cardiomegaly | 8552 | 21941 | 2564 | 5059 |

In order to validate the performance of the proposed method against the results of Irvin [9], one other pathology that is reported in their work was extracted and used to train a separate network. For this purpose, cardiomegaly was selected, resulting in 11 116, and 27 000 cases with positive and negative labels, respectively.
The code is publicly available at https://github.com/artinmajdi/chest_xray.

## 3. EXPERIMENTAL RESULTS

Fig. 3 shows two cases of patients with a pulmonary nodule and cardiomegaly. The location of each abnormality is overlaid onto the respective CXR image. As described in Sect. 2, the original images are heavily downsampled which makes the diagnosis of small nodules nearly impossible for a human expert. Even though the nodule image shown in Fig. 3 is reduced from 2828 × 2320 to 320 × 320, the proposed method is able to correctly classify the lung nodule. However, the proposed method will likely fail to detect nodules with sizes smaller than 10 × 10 voxels in the original image due to downsampling.

Table 2 shows the AUC per pathology for the proposed method and three existing methods. In order to be consistent, the value shown for Irvin [9] is taken from the case with "ignoring the uncertain labels" as reported in their work. We can see an increase in AUC in the proposed method with comparison to the existing methods for both pathologies.

Table 2: AUC FOR LUNG NODULE AND CARDIOMEGALY

|  | Proposed | Irvin [9] | Wang[11] | Yao [17] |
| --- | --- | --- | --- | --- |
| Pulmonary nodule | 0.73 | - | 0.67 | 0.72 |
| Cardiomegaly | 0.92 | 0.83 | 0.80 | 0.90 |

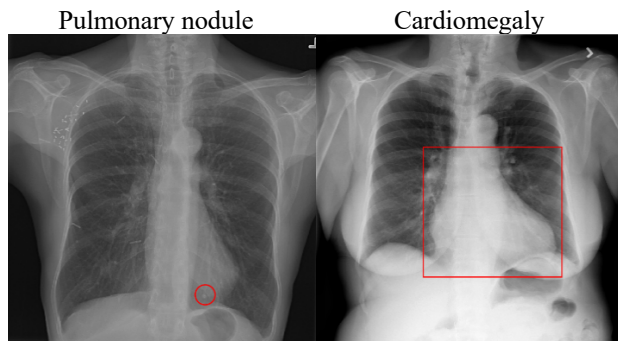

Figure 3: Examples of a pulmonary nodule (lung lesion) and cardiomegaly. The area encompassing the abnormalities is shown in red overlay.

Fig. 4 shows the ROC curves for each pathology and each label (presence or absence of that pathology). The ROC curve is measured by varying the discrimination threshold that is used to binarize the output probability maps. We can see a high (0.92) AUC for cardiomegaly and a moderate (0.73) AUC for pulmonary nodule classification. The lower AUC in the latter is likely due to the vast class imbalance (only 12.7% of the extracted dataset consisted of positive labels) and complexity of detecting cases with nodules (due to difficulty in detecting nodules in the vicinity of certain structures such as bone, and also the small size of nodules).

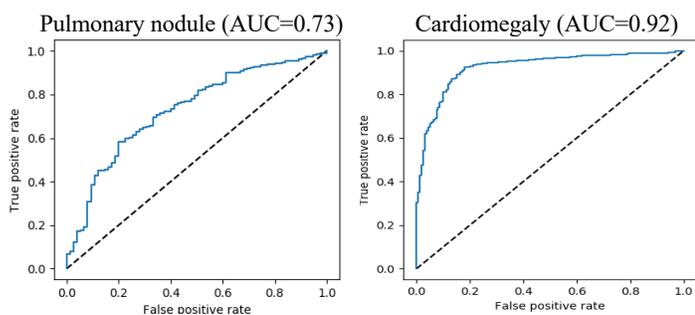

Figure 4: ROC curves of proposed algorithm for two pathologies (nodule (lesion) and cardiomegaly).

## 4. CONCLUSION

We have proposed a deep learning network for classification of pulmonary nodules and cardiomegaly. Based on ROC analysis, the proposed approach has successfully outperformed the existing methods. Specifically, the proposed network had a high AUC (0.91) for cardiomegaly classification, and a moderate AUC (0.73) for nodule classification.